\documentclass[lettersize,journal]{IEEEtran}
\usepackage{amsmath,amsfonts}
\pdfoutput=1
\usepackage{algorithmic}
\usepackage{algorithm}
\usepackage{array}
\usepackage[caption=false,font=normalsize,labelfont=sf,textfont=sf]{subfig}
\usepackage{textcomp}
\usepackage{stfloats}
\usepackage{url}
\usepackage{verbatim}
\usepackage{graphicx}
\usepackage{cite}
\begin{document}

\title{Bridging BCI and Communications: A MIMO Framework for EEG-to-ECoG Wireless Channel Modeling}

	\author{
		Jiaheng Wang, Zhenyu Wang, Tianheng Xu, Yuan Si, Ang Li, Ting Zhou, Xi Zhao, and Honglin Hu, Senior Member, IEEE
		\thanks{This paragraph of the first footnote will contain the date on 
			which you submitted your paper for review. }
		\thanks{Honglin Hu is with   Shanghai Advanced Research Institute, Shanghai 200100, China,  and also with ShanghaiTech University, Shanghai 200100, China (e-mail: huhl@sari.ac.cn). }
		\thanks{Jiaheng Wang, Zhenyu Wang, Tianheng Xu, Ang Li and Yuan Si are with   Shanghai Advanced Research Institute, Shanghai 200100, China (e-mail: wangjh@sari.ac.cn; wangzhenyu@sari.ac.cn; xuth@sari.ac.cn; lia@sari.ac.cn; siy@sari.ac.cn).}
		\thanks{Ting Zhou and Xi Zhao are with Shanghai University, Shanghai 201800, China  (e-mail: ting.zhou@mail.sim.ac.cn; zhaoxi2018@sari.ac.cn).}}
	\maketitle

% The paper headers
\markboth{Journal of \LaTeX\ Class Files,~Vol.~14, No.~8, August~2021}%
{Shell \MakeLowercase{\textit{et al.}}: A Sample Article Using IEEEtran.cls for IEEE Journals}

\maketitle

This work has been submitted to the IEEE for possible publication. Copyright may be transferred without notice, after which this version may no longer be accessible.

\begin{abstract}

% Bridging brain and external devices, Brain-Computer Interfaces (BCIs)  has attracted extensive research attention. 
As a method to connect human brain and external devices, Brain-computer interfaces (BCIs) are receiving extensive research attention.
Recently, the integration of communication theory with BCI has emerged as a popular  trend, offering  potential to enhance system performance and shape next-generation communications.
 A key challenge in this field is modeling the brain wireless communication channel between intracranial electrocorticography (ECoG) emitting neurons and extracranial electroencephalography (EEG)  receiving electrodes. However, the complex physiology of brain challenges the application of traditional channel modeling methods, leaving relevant research in its infancy.
To address  this gap, we propose a frequency-division multiple-input multiple-output (MIMO) estimation framework leveraging simultaneous macaque EEG and ECoG recordings, while employing neurophysiology-informed regularization to suppress noise interference. This approach reveals profound similarities between neural signal propagation and multi-antenna communication systems. Experimental results show improved estimation accuracy over conventional methods while highlighting a trade-off between frequency resolution and temporal stability determined by signal duration. This work  establish a conceptual bridge between neural interfacing and communication theory,  accelerating synergistic developments in both fields.

% This pioneering work is set to advance the integration of BCI research with wireless communication theory.

\end{abstract}

\begin{IEEEkeywords}
Brain–computer interface, wireless channel modeling, MIMO, brain-computer communication, biomedical communication.
\end{IEEEkeywords}

\section{Introduction}

\IEEEPARstart{B}{rain-computer } interface (BCI), serving as a direct communication bridge between the brain and external systems, has garnered sustained attention. Recent studies suggest  that BCI constitutes a typical  communication system \cite{hu2024survey}\cite{moioli2021neurosciences}, where 
synchronously activated neurons function as large-scale transmitting antennas, continuously emitting  electrocorticographic (ECoG)  signals encoding neural activity information. Following transmission through multilayered biological media, these signals are captured by  electroencephalographic (EEG) electrodes on the scalp and ultimately decoded by external devices.
The corresponding emerging domain, named BCI-Inspired Communications\cite{hu2022guest}, has also seen early exploration of its key concepts, including wireless-based BCI \cite{simeral2021home}, communication modulation in  BCI\cite{hu2024survey}, brain-to-brain communications \cite{pais2015building}, and brain access to 6G network\cite{moioli2021neurosciences}\cite{saad2019vision}. 
The convergence of BCI and modern communication demonstrates transformative potential to attain ultra-low latency, augmented precision, multifunction and high reliability\cite{hu2024survey}, which may become pivotal for next-generation BCI technology\cite{gao2021interface}. 
% might help BCI supersede conventional interfaces through intuitive and seamless neuro-environmental interaction.

In BCI-Inspired Communications, accurate wireless channel modeling—capturing the pathways from ECoG emitting neurons to EEG receiving electrodes—serves as the critical prerequisite for communication technology applications \cite{hu2024survey}. 
The modeling reveals how neural signals propagate may and offers basic guidance for techniques like brain-associated modulation  and adaptive decoding\cite{ramadan2017brain}, potentially pushing BCI performance beyond current threshold. 
However,   although conventional wireless channel modeling benefit from  mature mathematical formalisms, the unique biological  disparities characterizing neural transmission pathways \cite{johari2023noise}  have made the  methodologies transfer considerably difficult, resulting in pronounced  vacancy of dedicated brain channel explorations.  This research gap has relegated BCI-Inspired Communications to preliminary investigations marked by  loose coupling between BCI and communication frameworks.

\begin{figure}[!t]
\centering
\includegraphics[width=0.85\columnwidth]{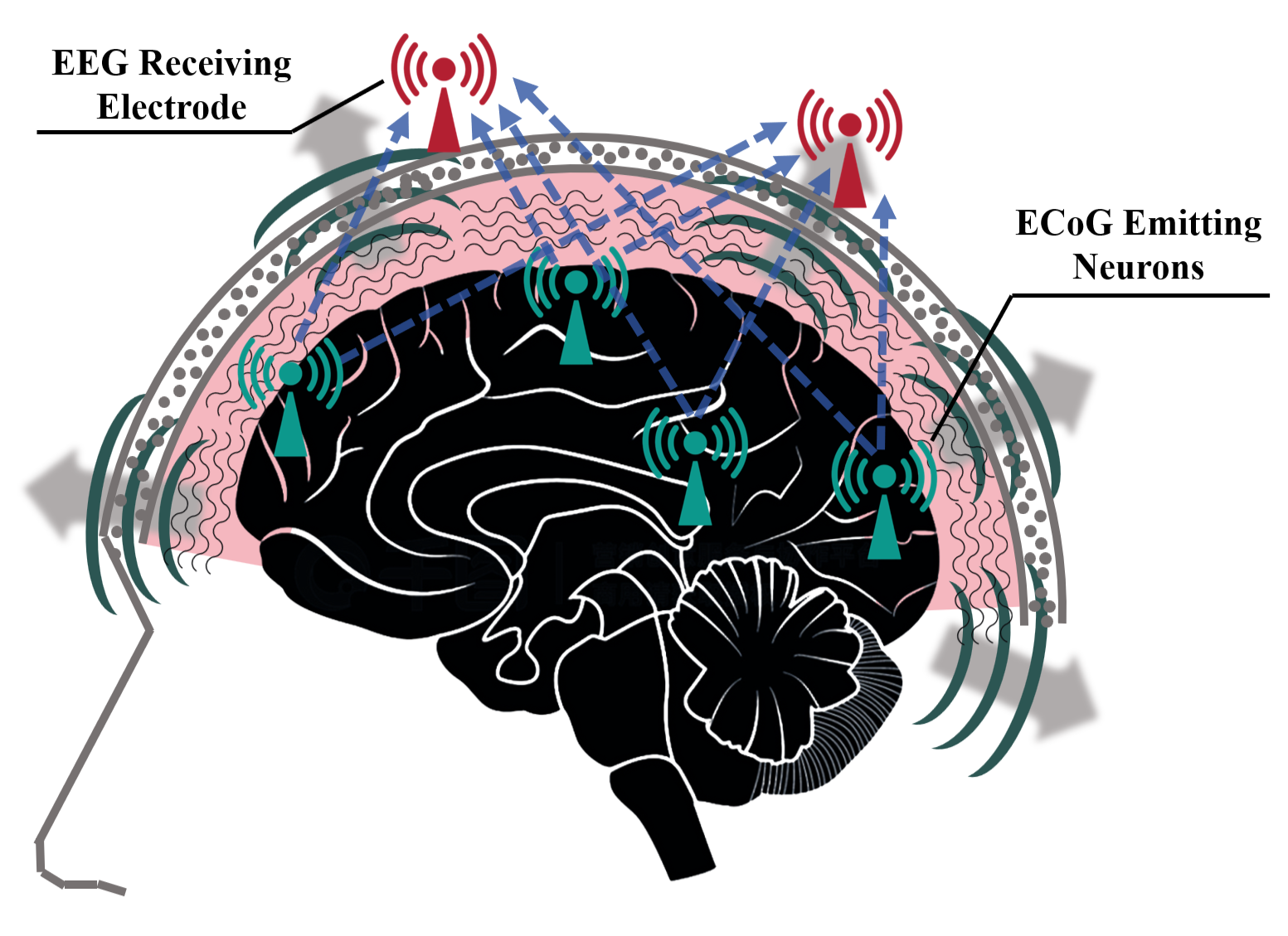}
\caption{The conceptual diagram of brain channel.}
\label{fig_1}
\end{figure}

 To address the research gap, we present a frequency-division multiple-input multiple-output (MIMO) estimation framework, FD-MIMO, derived from synchronized  EEG and ECoG recordings in non-human primates. Our approach identifies deep similarities between neural signal propagation pathways and multi-antenna communication systems,
 thereby extending wireless channel modeling to biological neural systems. 
  Additionally, traditional channel estimation methods fail to leverage brain physiological priors and address the complex noise  in neural signal propagation, resulting  overfitting and degraded estimation performance.  
 For this, we formulate physiological  properties as dual regularization—spatial smoothness and temporal continuity—to guide the channel estimation towards convergence  aligning with actual physiological mechanisms. 
Furthermore, we identify the crucial temporal-spectral trade-off  in signal segmentation, with the consistency between analytical predictions and experimental observations further confirming the universe applicability of communication theory in neural information transmission.  Ultimately, we achieve low-error brain channel estimation, which is expected to establish foundational insights for pioneering BCI-Inspired Communications.

\section{System Model}

\subsection{Frequency-division MIMO Estimation Framework}

\begin{figure*}[!t]
\centering
\includegraphics[width=1.5\columnwidth]{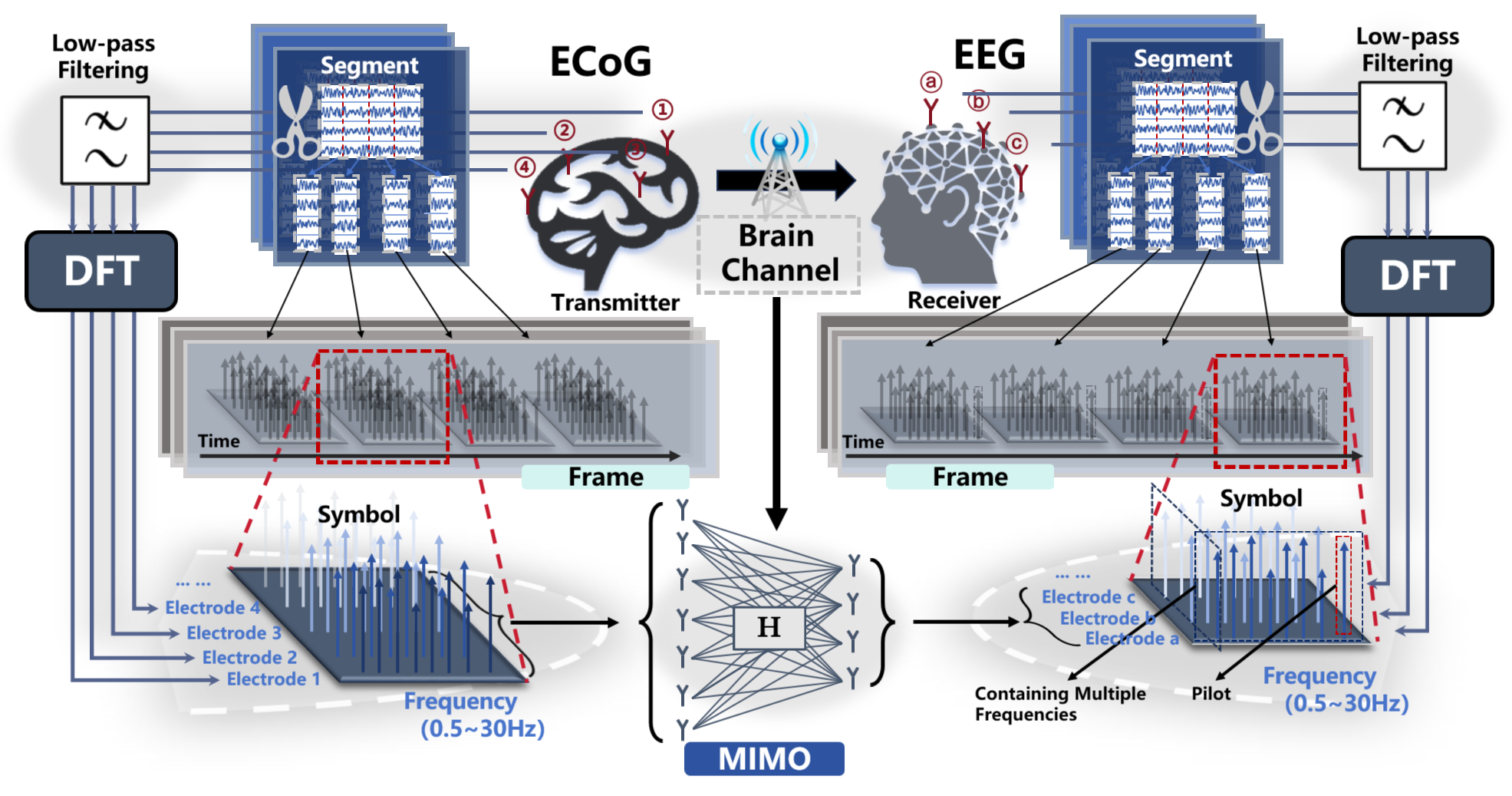}
\caption{The illustration of FD-MIMO.}
\label{fig_2}
\end{figure*}

The brain wireless channel fundamentally constitutes a MIMO communication system\cite{marzetta2016fundamentals}, as illustrated in Fig. \ref{fig_1}. In Fig. \ref{fig_1}, ECoG electrode arrays placed inside the skull form  signal sources, while  EEG sensors (red) on the scalp compose the reception array. Dashed blue lines represent the paths that signals take through the head encompassing  biological media such as cerebral cortex, meninges, skull, and scalp, etc., whose heterogeneous  dielectric properties induce attenuation, distortion, and filtering on the signal.

This motivates FD-MIMO, as seen in Fig. \ref{fig_2}. Firstly,  ECoG and EEG are uniformly segmented into $G$ symbols in time domain and   bandpass filtered by Butterworth within 0.3–400 Hz. Each symbol comprises $L$  sample points. Let $N$ denote the number of EEG electrodes, $P$ the number of ECoG electrodes, and $T$ the total sampling points ($T=G \cdot L$). The raw time-domain signals for the $n$-th ECoG and EEG symbols are then expressed as:
\begin{equation}
\mathbf{S}_{\mathrm{ECoG}}^{(n)} \in \mathbb{R}^{P \times L}, \quad 
\mathbf{S}_{\mathrm{EEG}}^{(n)} \in \mathbb{R}^{N \times L}.
\end{equation}

All symbols are grouped into $K$ frames, and  each  frame contains $M$ symbols  (thus $G = M \cdot K$). The discrete Fourier transform (DFT) is applied to each symbol, where the frequency resolution $\Delta f$ is determined by
$\Delta f = \frac{F_s}{L}$,
with $F_s$ denoting the sampling rate of time-domain EEG and ECoG.

We retain frequency components corresponding to $f = 0, \Delta f, 2\Delta f, \dots, 30\,\mathrm{Hz}$, representing the set of physiologically relevant neural oscillation frequencies (denoted by $\mathcal{F}$). The frequency-domain ECoG and EEG at frequency point $f$ of the $k$-th frame  are  denoted as $\mathbf{X}_k^{(f)} \in \mathbb{C}^{P \times M}$ and $\mathbf{Y}_k^{(f)} \in \mathbb{C}^{N \times M}$, respectively. This can be expressed as:
% \begin{equation}
% \begin{aligned}
% \mathbf{X}_k^{(f)} = \mathrm{DFT}\left( \mathbf{S}_{\mathrm{ECoG}}^{(k)} \right) \Bigr|_{f \in \mathcal{F}}, \quad
% \mathbf{Y}_k^{(f)} = \mathrm{DFT}\left( \mathbf{S}_{\mathrm{EEG}}^{(k)} \right) \Bigr|_{f \in \mathcal{F}}.
% \end{aligned}
% \end{equation}

\begin{equation}
\scalebox{0.75}{$\displaystyle
\begin{aligned}
\mathbf{X}_k^{(f)} &= \left[
\mathrm{DFT}\left( \mathbf{S}_{\mathrm{ECoG}}^{((k-1)M+1)} \right),\,
\mathrm{DFT}\left( \mathbf{S}_{\mathrm{ECoG}}^{((k-1)M+2)} \right),\,
\cdots,\,
\mathrm{DFT}\left( \mathbf{S}_{\mathrm{ECoG}}^{((k-1)M+M)} \right)
\right]\Big|_{f \in \mathcal{F}}, \\[1mm]
\mathbf{Y}_k^{(f)} &= \left[
\mathrm{DFT}\left( \mathbf{S}_{\mathrm{EEG}}^{((k-1)M+1)} \right),\,
\mathrm{DFT}\left( \mathbf{S}_{\mathrm{EEG}}^{((k-1)M+2)} \right),\,
\cdots,\,
\mathrm{DFT}\left( \mathbf{S}_{\mathrm{EEG}}^{((k-1)M+M)} \right)
\right]\Big|_{f \in \mathcal{F}}.
\end{aligned}.$}
\end{equation}

The electrophysiological signal transmission is modeled as:
$\mathbf{Y}_k^{(f)} = \mathbf{H}_k^{(f)}\mathbf{X}_k^{(f)} + \mathbf{N}$,
where   $\mathbf{H}_k^{(f)} \in \mathbb{C}^{N \times P}$ denotes MIMO channel matrix, and $\mathbf{N}$ represents noise.   $\mathbf{H}_k^{(f)}$ is estimated independently for each frequency $f$ across $k$ frames, through channel estimation methods like Least Squares (LS).

\subsection{Neurophysiologically  Motivated spatial and Temporal Constraints}

Compared to conventional communication channels, brain channel presents a significant challenge of physiological artifacts and environmental inference\cite{kalita2024aneeg}\cite{liu2022research}. Consequently, the performance of traditional channel estimation algorithms like LS  and Minimum Mean Square Error (MMSE)  is compromised. 
To address this, we formalize the physiological nature of neural signal propagation as math regularization. The approach leads to the channel estimation that is consistent with physiological priors, enabling robust and interpretable brain channel modeling in high noise.

Specifically, two brain-inspired regularizations are derived.
The first is spatial smoothness constraint. Adjacent electrodes demonstrate correlated signal propagation characteristics owing to overlapping tissue conduction pathways. We consequently impose similarity constraints between channel response row vectors $\mathbf{H}(i,:)$ and $\mathbf{H}(j,:)$ for neighboring electrodes $i$ and $j$.

The second is temporal continuity constraint. Cranial biophysical characters (e.g., skull conductivity, scalp thickness, and the spatial relationships between these tissues) maintain temporal stability over brief temporal intervals. We therefore regularize the variation between temporally successive channel matrices $\mathbf{H}_k^{(f)}$ and $\mathbf{H}_{k-1}^{(f)}$ to enforce temporal consistency with neural signal dynamics, while mitigating distortions between frames caused by noise. 
Accordingly,  \(\mathbf{H}_k^{(f)}\) partially relies on the previous estimated \(\mathbf{H}_{k-1}^{(f)}\).

The method named STARE (Spatial-Temporal Adaptive Regularized Estimation) is formulated through the objective function:

\begin{align}
&\min_{\mathbf{H}_k^{(f)}} J(\mathbf{H}_k) = \min_{\mathbf{H}_k^{(f)}} \;\;
 \underbrace{\frac{1}{2}\,\|\mathbf{Y}_k^{(f)} - \mathbf{H}_k^{(f)} \,\mathbf{X}_k^{(f)}\|_F^2}_{\text{Data Authenticity  Term}}
\;\\+\;
&\mu \underbrace{\sum_{(i,j)\in \mathcal{E}} \bigl\|\mathbf{H}_k^{(f)}(i,:) - \mathbf{H}_k^{(f)}(j,:)\bigr\|_F^2}_{\text{Spatial Smoothness Constraint}} \nonumber 
 +
\nu \underbrace{\bigl\|\mathbf{H}_k^{(f)} - \mathbf{H}_{k+1}^{(f)}\bigr\|_F^2}_{\text{Temporal Continuity Constraint}},
\label{objective}
\end{align}
where $\mathcal{E}$ denotes the set of all neighboring electrode pairs, with $\mu$ and $\nu$ as regularization parameters.
 $J(\mathbf{H}_k)$ consists of three  terms. Firstly, data authenticity  term ensures channel estimation accuracy, enforcing consistency with actual transmission observations by minimizing the Frobenius norm between $\mathbf{Y}_k^{(f)}$ and $\mathbf{H}_k^{(f)}\mathbf{X}_k^{(f)}$.

Secondly, the spatial smoothness constraint is designed based on the physical topology of the electrode array: For each spatially adjacent electrode pair $(i,j) \in \mathcal{E}$, we regularize the similarity between corresponding channel response  $\mathbf{H}(i,:)$ and $\mathbf{H}(j,:)$, for suppressing noise-induced spatial discontinuities. To facilitate derivation of closed-form solutions for $\mathbf{H}_{k}^{(f)}$, treating the physical adjacency relationships of all electrodes as an undirected graph, we reformulate the expression:
\begin{equation}
\scalebox{0.89}{$\displaystyle
\sum_{(i,j)\in \mathcal{E}} \bigl\|\mathbf{H}_k^{(f)}(i,:) - \mathbf{H}_k^{(f)}(j,:)\bigr\|_F^2 = 2\operatorname{Tr}\!\bigl((\mathbf{H}_k^{(f)})^{\top}\,\mathbf{L}^{\top}\mathbf{L}\,\mathbf{H}_k^{(f)}\bigr),
$}
\end{equation}
where the graph Laplacian matrix $\mathbf{L} \in \mathbb{R}^{P \times P} = \mathbf{D} - \mathbf{A}$ derives from the degree matrix $\mathbf{D}$ and adjacency matrix $\mathbf{A}$. The diagonal degree matrix $\mathbf{D}$ contains vertex degrees $D_{ii}$ representing the number of adjacent electrodes, while the symmetric binary adjacency matrix $\mathbf{A}$ satisfies $A_{ij}=1$ iff electrodes $i$ and $j$ are adjacent.

Thirdly, the temporal continuity constraint leverages the previous channel estimate $\mathbf{H}_{k-1}^{(f)}$ to enable temporal information propagation across frames.

\subsection{Solving Optimization Problem with  ADMM}

While the solution of the optimization problem  can be obtained by taking $\frac{\partial J(\mathbf{H}_k^{(f)})}{\partial \mathbf{H}_k^{(f)}}$ and setting it to zero, this approach cannot yield a closed-form solution for $\mathbf{H}_{k}^{(f)}$ due to constraints coupling. We therefore introduce an auxiliary variable $\mathbf{G}_k$ to decouple spatial and temporal constraints:
\begin{equation}
\scalebox{0.9}{$\displaystyle
\begin{aligned}
\min_{\mathbf{H}_k, \mathbf{G}_k} \quad & \frac{1}{2} \|\mathbf{Y}_k - \mathbf{H}_k \mathbf{X}_k\|_F^2 +\nu \|\mathbf{H}_k - \mathbf{H}_{k-1}\|_F^2 \\ &+ 2\mu \, \mathrm{Tr}(\mathbf{G}_k^\top \mathbf{L}^\top \mathbf{L} \mathbf{G}_k)   \quad \quad 
\text{s.t.} \quad  \mathbf{G}_k = \mathbf{H}_k.
\end{aligned}$}
\end{equation}
This is a two-variable, single-constraint optimization.
Thus, we employ the Alternating Direction Method of Multipliers (ADMM) \cite{han2022survey} to  update $\mathbf{H}_k$ and $\mathbf{G}_k$. ADMM is an iterative algorithm. Given  the fixed frequency index  $f$, we omit this superscript and replace it with the iteration index $t$  for notational clarity. The augmented Lagrangian formulation becomes:

\begin{equation}
\scalebox{0.9}{$\displaystyle
\begin{split}
\mathcal{L}_k(\mathbf{H}_k, \mathbf{G}_k, \mathbf{\Lambda}_k) &= \frac{1}{2}\|\mathbf{Y}_k - \mathbf{H}_k \mathbf{X}_k\|_F^2 + 2\mu \mathrm{Tr}((\mathbf{G}_k)^{\top}\mathbf{L}^{\top} \mathbf{L} \mathbf{G}_k) \\+ \nu \|\mathbf{H}_k - \mathbf{H}_{k-1}\|_F^2 
&+\mathrm{Tr}(\mathbf{\Lambda}_k^\top (\mathbf{G}_k - \mathbf{H}_k))+ \frac{\rho}{2} \|\mathbf{G}_k - \mathbf{H}_k \|_F^2,
\end{split}$}
\end{equation}
where $\mathbf{\Lambda}_k$ denotes dual variables and $\rho > 0$ represents a penalty parameter.

The algorithm initialization leverages temporal continuity prior by setting $\mathbf{H}_{k}^{(0)} = \mathbf{H}_{k-1}$,  $\mathbf{G}_k^{(0)} = \mathbf{H}_{k-1}$ and $\mathbf{\Lambda}_k^{(0)} = \mathbf{0}$. At iteration $t$, the update steps proceed as follows.

\noindent\textit{(1)}  Update $\mathbf{H}_k$, with fixed $\mathbf{G}_k^{(t)}$ and $\mathbf{\Lambda}_k^{(t)}$:
\begin{equation}
\scalebox{0.9}{$\displaystyle
\begin{split}
\mathbf{H}_k^{(t+1)} = \arg\min_{\mathbf{H}_k}& \frac{1}{2}\|\mathbf{Y}_k - \mathbf{H}_k \mathbf{X}_k\|_F^2 + \nu \|\mathbf{H}_k - \mathbf{H}_{k-1}\|_F^2  \\ &+\frac{\rho}{2} \|\mathbf{G}_k^{(t)} - \mathbf{H}_k + \mathbf{\Lambda}_k^{(t)}\|_F^2,\end{split}$}
\end{equation}
yielding the closed-form solution:
\begin{equation}
\scalebox{0.83}{$\displaystyle
\mathbf{H}_k^{(t+1)} = \left( \mathbf{Y}_k \mathbf{X}_k^H + 2\nu \mathbf{H}_{k-1} + \rho (\mathbf{G}_k^{(t)} + \mathbf{\Lambda}_k^{(t)}) \right) \left( \mathbf{X}_k \mathbf{X}_k^H + 2\nu \mathbf{I} + \rho \mathbf{I} \right)^{-1}.
$}
\end{equation}

\noindent\textit{(2)} Update $\mathbf{G}_k$, with fixed $\mathbf{H}_k^{(t+1)}$ and $\mathbf{\Lambda}_k^{(t)}$:
\begin{equation}
\scalebox{0.9}{$\displaystyle
\mathbf{G}_k^{(t+1)} = \arg\min_{\mathbf{G}_k} 2\mu \mathrm{Tr}(\mathbf{G}_k^{\top}\mathbf{L}^{\top} \mathbf{L} \mathbf{G}_k) + \frac{\rho}{2} \|\mathbf{G}_k - \mathbf{H}_k^{(t+1)} + \mathbf{\Lambda}_k^{(t)}\|_F^2.$}
\end{equation}
The closed-form solution is:
\begin{equation}
\scalebox{1}{$\displaystyle
\mathbf{G}_k^{(t+1)}  = (4\mu \mathbf{L}^\top \mathbf{L} + \rho \mathbf{I})^{-1} \rho (\mathbf{H}_k^{(t+1)} - \mathbf{\Lambda}_k^{(t)}).$}
\end{equation}

\noindent\textit{(3)} Update $\mathbf{\Lambda}_k^{(t)}$:
\begin{equation}
\scalebox{1}{$\displaystyle
\mathbf{\Lambda}_k^{(t+1)} = \mathbf{\Lambda}_k^{(t)} + \rho (\mathbf{G}_k^{(t+1)} - \mathbf{H}_k^{(t+1)}).$}
\end{equation}

Repeat these three steps until the maximum number of count $t_{\max}$ is reached.

\begin{figure*}[!h]
\centering
\includegraphics[width=1.3\columnwidth]{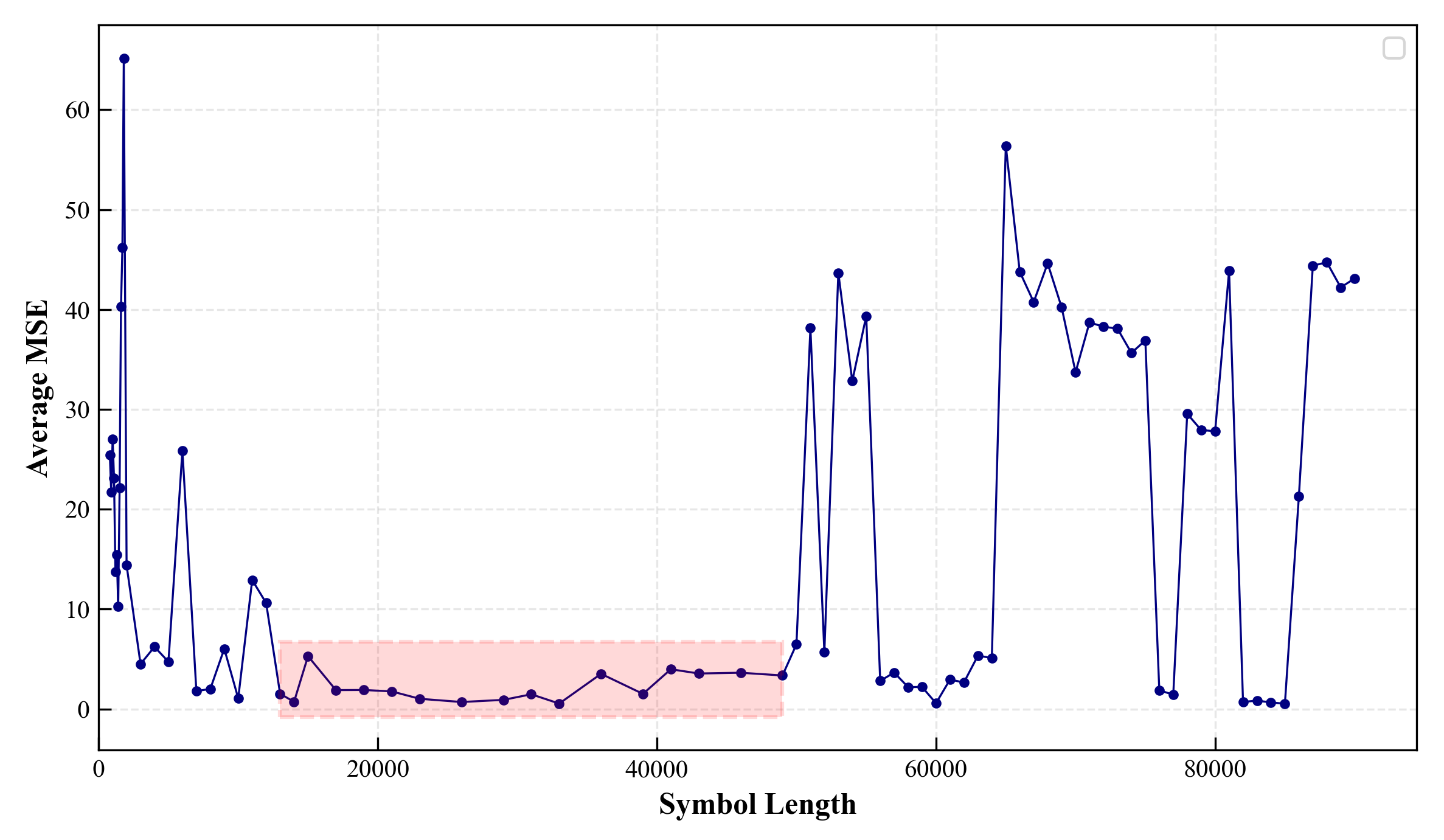}
\caption{The relationship between the symbol length and the $\text{MSE}_{Avg}$.
}
\label{fig_3}
\end{figure*}

\begin{figure}[!b]
\centering
\includegraphics[width=0.9\columnwidth]{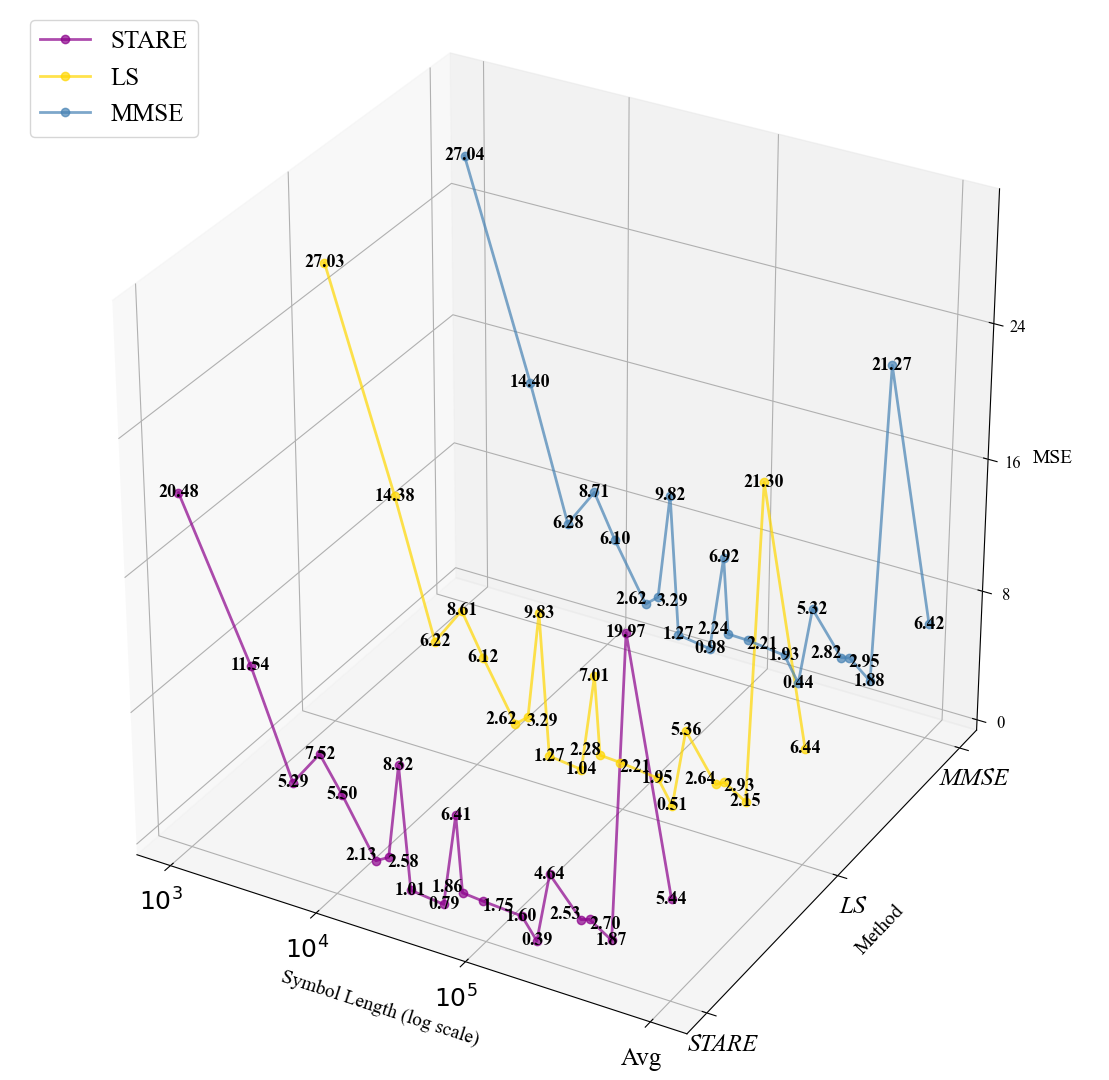}
\caption{The $\text{MSE}_{Avg}$ at the corresponding symbol length for three methods: STARE, LS, and MMSE. At the far end of the axis, the average MSE across all symbol lengths is presented: 5.44 for STARE, compared to 6.44 for LS and 6.42 for MMSE.
}
\label{fig_4}
\end{figure}

Finally, The full-band channel response is synthesized through  combining (in the frequency domain) the responses from all frequency points:
\begin{equation}
	\scalebox{0.9}{
\scalebox{1}{$\displaystyle
\mathbf{H}_k = \bigoplus_{f \in \mathcal{F}} \mathbf{H}_k^{(f)} \in \mathbb{C}^{N \times P \times |\mathcal{F}|}.$}}
\end{equation}

\section{Experimental Results}

\subsection{Evaluation Metric and Dataset}
The metric quantifying brain channel estimation errors requires consideration. While mean squared error (MSE) serves as a standard metric for evaluating channel modeling, the absence of ground-truth brain channel parameters precludes direct computation of MSE between estimated and actual channels. To circumvent this limitation, first, our approach generates simulated EEG signals through $\mathbf{\hat{Y}}_k^{(f)} = \mathbf{\hat{H}}_k^{(f)}\mathbf{X}_k^{(f)}$ using the estimated channel matrix $\mathbf{\hat{H}}_k^{(f)}$. The reconstruction error on frequency $f$ is then calculated as 
\begin{equation}
\text{MSE} = \|\mathbf{Y}_{k\;\mathbf{true}}^{(f)} - \mathbf{\hat{Y}}_k^{(f)}\|_F^2,
\label{mse}
\end{equation}
where $\mathbf{Y}_{k\;\mathbf{true}}^{(f)}$ is recorded EEG.
That is, brain channel estimation algorithms achieving greater similarity (i.e., reduced MSE) between  simulated EEG  and real EEG recordings exhibit valid estimation performance.

Experimental validation of brain channel modeling necessitates simultaneous EEG and ECoG recordings, which is scarce for human. However, non-human primate  present a promising alternative, specifically macaque. In terms of skull anatomy \cite{vajhi2023ct}, cortical tissue electrophysiological properties  \cite{herrera2022resolving}  
%brain region functions \cite{casimo2017interspecies}\cite{lu2023connectivity}
 and neural signal origins\cite{cohen2009origin}\cite{oosugi2017new}, macaques closely resemble humans. 
 Consequently, macaque  experiments  often serve as  valuable preliminary endeavors for subsequent human studies   \cite{oosugi2017new}. Notably, the Multidimensional Recording (MDR) dataset (\url{http://neurotycho.org}) provides the sole publicly available  whole-cortex simultaneous EEG-ECoG recordings from two adult macaque monkeys\cite{oosugi2017new}.
Prior to recording,  skin, skull, and dura mater of the macaque were carefully resected. A high-density 256 ECoG electrode array (3 mm diameter) was  deployed with 5 mm inter-electrode spacing across frontal, parietal, temporal,  occipital lobes and  medial wall  (electrodes 1--128 are on the left hemisphere, and 129--256 are on the right). Following ECoG electrode placement, the dura mater, skull, and skin were meticulously repositioned.
% All procedures adhered to RIKEN ethical committee (No. H22-2-202(4)) and recommendations of the Weatherall report, “the use of non-human primates in research”.
% Data acquisition occurred under two consciousness states (awake and anesthetized) using a primate chair apparatus with restrained limbs and visual occlusion.
 Both ECoG and EEG  were sampled at 1,000 Hz, with 17 EEG electrodes arranged according to the 10--20 system without Cz and Pz. 
 % This  dataset precisely satisfies the requirements for  brain channel estimation.
% While clinical epilepsy research provides limited intracranial recordings \cite{frauscher2018atlas}, such data typically offers restricted cortical coverage focused on seizure regions, inadequate for whole-cortex brain channel modeling. 

\subsection{Symbol Length Optimization in Brain Channel Estimation}
Segmenting ECoG and EEG into symbols constitutes a fundamental step for FD-MIMO.
The  symbol length $L$  critically influences estimation performance yet without established guidance. Here, we provide  theoretical analysis and empirical validation for this issue.

Given a fixed sampling rate $F_s$, we have:
\begin{equation}
\Delta f = \frac{F_s}{L}.
\end{equation}
As mentioned above, brain channel estimation operates discretely across individual frequency points. Frequency resolution $\Delta f$ is thus  defined as the minimum resolvable frequency interval between adjacent frequencies.
$\Delta f$ directly determines the capability of FD-MIMO to discriminate  adjacent spectral components of neural activity.

 Theoretically, insufficient  $L$ ($L \ll L_{\mathrm{opt}}$, where $ L_{\mathrm{opt}}$ is the optimal symbol length) degrade frequency resolution, inducing cross-frequency interference at target frequency $f$ and preventing effective separation of coexisting neural activities, thereby elevating estimation error.

Conversely,  excessively prolonged $L$ ($L \gg L_{\mathrm{opt}}$), while providing enhanced frequency resolution, may induce time-varying brain channel, wherein the channel characteristics evolve within a single symbol period. Besides, multiple physiological events could be encapsulated within an elongated symbol, resulting in intra-symbol spectral disparities.
The above mechanisms would compromise brain channel estimation. Therefore, an ideal  $L$ exists that balances  frequency resolution and temporal stationarity. This theoretical analysis provides critical guidance for  experiments.

In experiments, we  vary $L$. 
Fig. 3 delineates the relationship between symbol length and channel estimation performance. 
Here, for each symbol length, we calculate the MSE at every frequency point across all frames  using the  formula \eqref{mse}, and represent the performance by averaging all the computed MSE, denoted by $\text{MSE}_{Avg}$.
The U-shaped curve demonstrates three regimes: (1) short symbols ($L < 13000$ samples, $\Delta f > 0.077$ Hz) with escalated average MSE of 18.22 across the regime, (2) an optimal mid-range ($13000 \leq L \leq 49000$ samples, red shaded region) achieving minimum average MSE across the regime of 2.2, and (3) excessively long symbols ($L > 49000$) exhibiting severe performance collapse ( average MSE across the regime of  23.7). Notably, the global minimum $\text{MSE}_{Avg}$ = 0.55 occurs at $L_{\mathrm{opt}}=33000$, with secondary performance excellence ($\text{MSE}_{Avg}$ $<$ 1) observed at 14000, 26000, 29000, and 60000 samples.

Above all, 
the performance degrades at both excessively short and prolonged symbol length, while an intermediate length range yields optimal estimation accuracy.
The precise  alignment between empirical evidence and theoretical analyses substantiates the  applicability of information theory in  modeling neural signal transmissions, and validating the feasibility of employing communication engineering frameworks to analyze and optimize brain channel modeling.

\subsection{Comparison with MMSE and LS}

 Fig. 4's graphical table delineates three methodologies - MMSE, LS, and STARE - through their performance (denoted by $\text{MSE}_{Avg}$) trajectories  across exponentially scaled symbol length ($10^3$ to $10^5$), with  discrete dots representing  recorded performance. The average MSE across all symbol lengths is 5.44 for STARE, compared to 6.44 for LS and 6.42 for MMSE. Compared to the LS and MMSE, STARE achieved reductions in MSE of 15.4\% and 15.3\%, respectively. Across all individual $L$, our method consistently outperforms the others. The result substantiates the global and local superiority of STARE through effective physiological prior utilization and brain channel noise mitigation. Furthermore, all methodologies exhibit a three-stage variation in performance with respect to $L$ (decline-stabilization-ascent pattern of $\text{MSE}_{Avg}$), validating the dual-impact of symbol length on frequency resolution and temporal  dynamics.

% Cerebral channel estimation exhibits performance degradation at both extremely short and prolonged symbol durations, while achieving optimal accuracy within an intermediate duration range (13,000-49,000 samples). The quantitative agreement between neurophysiological measurements and communication-theoretic principles substantiates information theory's universal applicability in modeling neural signal transmission mechanisms, thereby validating the technical feasibility of employing communication engineering frameworks to analyze and optimize neuro-electromagnetic information pathways.

% The  U-curve with demonstrates minimum  MSE  within the intermediate duration window (13,000-49,000 samples, blue box, the average MSE across the box is 2.2), contrasting sharply with degraded performance in short ($L < 13,\!000$, $\Delta f > 0.077$) and long ($L > 49,\!000$) regimes where MSE escalates to 18.22 and 23.7 respectively. Notably, the optimal duration of $L = 33,\!000$ achieves peak estimation precision (MSE = 0.55), with secondary performance excellence (MSE < 1) observed at 14,000, 26,000, 29,000, and 60,000 samples.

\section{Conclusion}

% This study addresses the fundamental challenge of brain channel modeling in \emph{BCI-Inspired Communications} through three key contributions. First, we propose a frequency-division MIMO estimation framework incorporating spatiotemporal regularization derived from biophysical priors. This approach effectively mitigates complex neural noise interference, achieving 15.4\% and 15.3\% lower mean squared error compared to conventional LS and MMSE estimators, respectively. Second, our analysis reveals an intrinsic trade-off mechanism between spectral resolution and temporal stability in symbol duration selection, validating the universal applicability of communication principles in neural systems. Third, we establish theoretical foundations for physical-layer design in neurocommunication systems by unifying biological constraints with information theory. While current results demonstrate theoretical advances, future work must address three critical challenges: (1) developing human-specific brain channel models, (2) designing adaptive equalization techniques for non-stationary neural channels, and (3) creating neuro-compatible modulation-coding schemes. This research bridges a crucial gap between computational neuroscience and wireless communication theory, paving the way for practical implementations of brain-inspired communication systems.

In this letter, we focus on the primary challenge in BCI-Inspired Communications---brain channel modeling---by proposing  FD-MIMO and STARE. Employing spatial-temporal regularization derived from biophysical priors, our approach effectively mitigates complex neural noise interference, yielding  15.4\% and 15.3\% lower mean squared error compared to conventional LS and MMSE estimators, respectively. Moreover, we reveal the trade-off between frequency resolution and time-varying stability in symbol segmentation, thereby validating the  universal applicability of communication principles in neural systems and establishing theoretical guidance for the physical layer design of  BCI-Inspired Communications Systems. This study  bridges a critical gap between BCI and wireless communication theory. Future work will explore human-specific brain channel models, channel equalization tailored to brain channels, and neuro-compatible modulation-coding schemes, hoping to pave the way for practical implementations of BCI-Inspired Communication Systems.

\bibliographystyle{IEEEtran}

\bibliography{IEEEabrv,Reference}%第二个参数参数是你的bib文件的名字

\end{document}